\documentclass[letter]{aa}
\usepackage{graphicx}
\usepackage{setspace}
\usepackage{natbib}
\usepackage{longtable}
\usepackage{amssymb}
\usepackage[varg]{txfonts}
\bibpunct{(}{)}{;}{a}{}{,}

\usepackage[normalem]{ulem}

\begin{document}

\newcommand{\FeII}{[\ion{Fe}{ii}]}
\newcommand{\TiII}{[\ion{Ti}{ii}]}
\newcommand{\SII}{[\ion{S}{ii}]}
\newcommand{\OI}{[\ion{O}{i}]}
\newcommand{\OIp}{\ion{O}{i}}
\newcommand{\PII}{[\ion{P}{ii}]}
\newcommand{\NI}{[\ion{N}{i}]}
\newcommand{\NII}{[\ion{N}{ii}]}
\newcommand{\NIp}{\ion{N}{i}}
\newcommand{\NiII}{[\ion{Ni}{ii}]}
\newcommand{\CaIIp}{\ion{Ca}{ii}}
\newcommand{\PI}{[\ion{P}{i}]}
\newcommand{\CIp}{\ion{C}{i}}
\newcommand{\HeI}{\ion{He}{i}}
\newcommand{\MgIp}{\ion{Mg}{i}}
\newcommand{\MgIIp}{\ion{Mg}{ii}}
\newcommand{\NaI}{\ion{Na}{i}}
\newcommand{\HI}{\ion{H}{i}}
\newcommand{\brg}{Br$\gamma$}
\newcommand{\pab}{Pa$\beta$}

\newcommand{\macc}{$\dot{M}_{acc}$}
\newcommand{\lacc}{L$_{acc}$}
\newcommand{\lbol}{L$_{bol}$}
\newcommand{\mjet}{$\dot{M}_{jet}$}
\newcommand{\mh}{$\dot{M}_{H_2}$}
\newcommand{\Ne}{n$_e$}
\newcommand{\h}{H$_2$}
\newcommand{\kms}{km\,s$^{-1}$}
\newcommand{\um}{$\mu$m}
\newcommand{\lam}{$\lambda$}
\newcommand{\msyr}{M$_{\sun}$\,yr$^{-1}$}
\newcommand{\Av}{A$_V$}
\newcommand{\msun}{M$_{\sun}$}
\newcommand{\lsun}{L$_{\sun}$}
\newcommand{\rsun}{R$_{\sun}$}
\newcommand{\cm}{cm$^{-3}$}
\newcommand{\ergscm}{erg\,s$^{-1}$\,cm$^{-2}$}

\newcommand{\tw}{TW\,Hya}

\newcommand{\bet}{$\beta$}
\newcommand{\alfa}{$\alpha$}

\hyphenation{mo-le-cu-lar pre-vious e-vi-den-ce di-ffe-rent pa-ra-me-ters ex-ten-ding a-vai-la-ble excited GRA-VI-TY}

\title{The GRAVITY young stellar object survey.}
\subtitle{II. First spatially resolved observations of the CO bandhead emission in a high-mass YSO.}

\author{GRAVITY Collaboration(\thanks{GRAVITY is developed in a
collaboration by the Max Planck Institute for Extraterrestrial Physics,
LESIA of Paris Observatory and IPAG of Université Grenoble Alpes / CNRS,
the Max Planck Institute for Astronomy, the University of Cologne, the
Centro Multidisciplinar de Astrofisica Lisbon and Porto, and the European
Southern Observatory.}): A. Caratti o Garatti\inst{1,2,3},
R. Fedriani\inst{1,3,4}, R. Garcia Lopez\inst{1,2,3}, M. Koutoulaki\inst{1,3,5}, K. Perraut\inst{6},
H. Linz\inst{2}, W. Brandner\inst{2}, P. Garcia\inst{7,8,9}, L. Klarmann\inst{2}, T. Henning\inst{2}, L. Labadie\inst{10}, J. Sanchez-Bermudez\inst{2,11}, B. Lazareff\inst{6},
E.F. van Dishoeck\inst{12,14}, P. Caselli\inst{12}, P.T. de Zeeuw\inst{12,14}, A. Bik\inst{13}, M. Benisty\inst{6,9}, C. Dougados\inst{6}, T.P. Ray\inst{1}, A. Amorim\inst{8},
J.-P. Berger\inst{6}, Y. Cl\'enet\inst{15}, V. Coud\'e du Foresto\inst{15}, G. Duvert\inst{6}, A. Eckart\inst{10}, F. Eisenhauer\inst{12}, F. Gao\inst{12}, E. Gendron\inst{15}, R. Genzel\inst{12,16}, S. Gillessen\inst{12}, P. Gordo\inst{8}, 
L. Jocou\inst{6}, M. Horrobin\inst{10}, P. Kervella\inst{15}, S. Lacour\inst{15}, J.-B. Le Bouquin\inst{6}, P. L\'ena\inst{15},
R. Grellmann\inst{10}, T. Ott\inst{12}, T. Paumard\inst{15}, G. Perrin\inst{15}, G. Rousset\inst{15}, S. Scheithauer\inst{2}, J. Shangguan\inst{12}, J. Stadler\inst{12}, O. Straub\inst{12}, C. Straubmeier\inst{10},
E. Sturm\inst{12}, W.F. Thi\inst{12}, F.H. Vincent\inst{15}, F. Widmann\inst{12}}
\institute{Dublin Institute for Advanced Studies, 31 Fitzwilliam Place, D02\,XF86 Dublin, Ireland
\and
 Max Planck Institute for Astronomy, K\"{o}nigstuhl 17, Heidelberg, Germany, D-69117
\and 
 University College Dublin, School of Physics, Belfield, Dublin 4, Ireland
\and
Department of Space, Earth \& Environment, Chalmers University of Technology, SE-412 93 Gothenburg, Sweden
\and
European Southern Observatory, Karl-Schwarzschild-Str. 2, D-85748, Garching, Germany
\and
Univ. Grenoble Alpes, CNRS, IPAG, F-38000 Grenoble, France
\and
Universidade do Porto - Faculdade de Engenharia, Rua Dr. Roberto Frias, 4200-465 Porto, Portugal
\and
CENTRA, Instituto Superior Tecnico, Av. Rovisco Pais, 1049-001 Lisboa, Portugal
\and
European Southern Observatory, Casilla 19001, Santiago 19, Chile
\and
I. Physikalisches Institut, Universität zu Köln, Zülpicher Str. 77, 50937, Köln, Germany
\and
Instituto de Astronom\'ia, Universidad Nacional Aut\'onoma de M\'exico, Apdo. Postal 70264, Ciudad de M\'exico 04510, Mexico
\and
Max Planck Institute for Extraterrestrial Physics, Giessenbachstrasse, 85741 Garching bei M\"{u}nchen, Germany
\and
Department of Astronomy, Stockholm University, Oskar Klein Center, SE-106 91 Stockholm, Sweden
\and
Leiden Observatory, Leiden University, P.O. Box 9513, NL-2300 RA, Leiden, The Netherlands
\and
LESIA, Observatoire de Paris, PSL Research University, CNRS, Sorbonne Universit\'es, UPMC Univ. Paris 06, Univ. Paris Diderot, Sorbonne Paris Cit\'e, France
\and
Department of Physics, Le Conte Hall, University of California, Berkeley, CA 94720, USA\\
\email{alessio@cp.dias.ie}
              }

   \date{Received ; accepted }

\titlerunning{Spatially resolved CO bandhead emission in NGC\,2024\,IRS\,2}
\authorrunning{Caratti o Garatti A. et al.}

  \abstract
   {The inner regions of the discs of high-mass young stellar objects (HMYSOs) are still poorly known due to the small angular scales and the high visual extinction involved.}
   {We deploy near-infrared (NIR) spectro-interferometry to probe the inner gaseous disc in HMYSOs and investigate the origin and physical characteristics of the CO bandhead emission (2.3--2.4\,$\mu$m).}
   {We present the first GRAVITY/VLTI observations at high spectral ($\mathcal{R} $=4000) and spatial (mas) resolution of the CO overtone transitions in \object{NGC 2024 IRS\,2}.} 
   {The continuum emission is resolved in all baselines and is slightly asymmetric, displaying small closure phases ($\leq$8$\degr$). Our best ellipsoid model  
   provides a disc inclination of 34$\degr\pm$1$\degr$, a disc major axis position angle ($PA$) of 166$\degr\pm$1$\degr$, and a disc diameter of 3.99$\pm$0.09\,mas (or 1.69$\pm$0.04\,au, at a distance of 423\,pc). The small closure phase signals 
   in the continuum are modelled with a skewed rim, originating from a pure inclination effect. For the first time, our observations spatially and spectrally resolve the first four CO bandheads. 
   Changes in visibility, as well as differential and closure phases across the bandheads are detected.
   Both the size and geometry of the CO-emitting region are determined by fitting a bidimensional Gaussian to the continuum-compensated CO bandhead visibilities.
   The CO-emitting region has a diameter of 2.74$\pm^{0.08}_{0.07}$\,mas (1.16$\pm$0.03\,au), and is located in the inner gaseous disc, well within the dusty rim, with inclination and $PA$ 
   matching the dusty disc geometry, which indicates that both dusty and gaseous discs are coplanar. Physical and dynamical gas conditions are inferred by modelling the CO spectrum.
  Finally, we derive a direct measurement of the stellar mass of  $M_*\sim$14.7$^{+2}_{-3.6}$\,M$_\sun$ by combining our interferometric and spectral modelling results.
  } 
   {}


   \maketitle
%

\section{Introduction}

Accretion discs around high-mass young stellar objects ($M > 8$\,M$_\sun$; O and early B spectral types) are key for understanding 
how massive stars form. However, their structure and main physical properties are poorly known~\citep[see][and references therein]{beltran16}. 
In particular, the study of the inner gaseous disc (within a few astronomical units from the central object), namely where accretion and ejection take place, can clarify what mechanisms are
at play (e.g., accretion from funnels or through boundary layers, ejection through stellar or MHD disc-winds, etc.). This crucial region still remains elusive because of the typically large distance (kiloparsecs; kpc) to HMYSOs 
and their high visual extinction ($A_V \geq$50\,mag). Therefore, near-infrared (NIR) spectro-interferometry is required to achieve milli-arcsecond (mas) spatial resolution and to spectrally resolve the warm gas (few thousand K)
that traces such processes. The CO overtone transitions (or bandheads) in the $K$-band (between 2.29 and 2.5\,$\mu$m) have been successfully employed to investigate both the kinematics and physics
of the inner gaseous disc in HMYSOs~\citep{blum,bik04}. The modelling of the CO bandheads profiles at high spectral resolution ($R \geq$10\,000) suggests that such emission comes from 
warm ($T$=2000--5000\,K) and dense ($n>$10$^{11}$\,cm$^{-3}$) gas in Keplerian rotation, within a few astronomical units from the central sources~\citep{ilee}, relatively close to the dust sublimation radius. 
However, this region has not been spatially resolved yet, and therefore its location and position are still uncertain. 

Here, we present the first spatially and spectrally resolved observations of the CO overtone transitions in a HMYSO, namely in \object{NGC 2024 IRS\,2}, using GRAVITY/VLTI spectro-interferometry. 
\object{NGC 2024 IRS\,2} (hereafter IRS\,2) is a well-studied HMYSO~\citep[$M_* \sim$15\,M$_\sun$; SpT$\sim$B0; $M_{disc}\sim$0.04\,M$_\sun$;][]{lenorzer,nisini94,mann}. 
Located in \object{NGC\,2024} within the \object{Orion\,B} complex at a distance of 423$\pm$15\,pc~\citep{2024distance}, IRS\,2 is very bright in the $K$-band~\citep[$K_s=4.585$\,mag;][]{2MASS}
and its spectrum shows strong CO bandheads~\citep{chandler,lenorzer}, making it an excellent test case for probing the origin of the NIR CO emission in HMYSOs.

\section{Observations and data reduction}

\begin{table*}[t]
\tiny
\caption{\label{tab:obslog} Observation log of the VLTI GRAVITY+UT high-resolution ($R$~$\sim$4000) observations of NGC\,2024\,IRS\,2. }
\centering
\vspace{0.1cm}
\begin{tabular}{@{}c c c c c c | c c }
\hline \hline
Date UT & Tot. Int. & DIT\tablefootmark{a} & NDIT\tablefootmark{b} & Proj. baselines & PA\tablefootmark{c} & Calibrator & UD diameter\tablefootmark{d}\\ 
yyyy-mm-dd hh:mm & [s] & [s] & [s] &  [m] & [$\circ$] & &[mas]\\  \hline 
2019-01-21 05:04& 900 & 30 & 10 &  45, 45, 56, 72, 101, 119 & 298, 46, 36, 82, 40, 62 & HD\,31464, HD\,37491 & 0.186$\pm$0.004, 0.464$\pm$0.018\\ 
2019-01-21 05:27& 900 & 30 & 10 &  44, 56, 100 & 45, 36, 40 & HD\,31464, HD\,37491 & 0.186$\pm$0.004, 0.464$\pm$0.018\\ 
2019-01-21 06:00& 900 & 30 & 10 &  34, 43, 55, 56, 98, 103 & 309, 43, 35, 80, 39, 53 & HD\,31464, HD\,37491 & 0.186$\pm$0.004, 0.464$\pm$0.018\\ 

\hline
\end{tabular}
\tablefoot{
\tablefoottext{a}{Detector integration time per interferogram.}
\tablefoottext{b}{Number of interferograms.}
\tablefoottext{c}{Baseline position angle ($PA$, from N to E) from the shortest to longest baseline.}
\tablefoottext{d}{The calibrator uniform-disc (UD) diameter (\textit{K} band) was taken from \cite{calibrators}
.}
}
\end{table*}

IRS\,2 was observed with GRAVITY/VLTI~\citep{GRAVITY} in the $K$-band (1.95--2.5\,$\mu$m) on 21$^{}$  January 2019 using the four 8\,m Unit Telescopes (UTs). 
The target was observed in single-field mode.
As both target and nearby stars are not visible in the optical, we used the IR wavefront sensing 
system CIAO~\citep[Coud\'e Infrared Adaptive Optics;][]{CIAO} guiding  off-axis on the nearby (4\arcsec.8) NIR star IRS\,2b. 
Three sets of data (with total integration time of 900\,s each) were acquired. Only UT1-2-3 data are present in the second dataset due to a technical failure at UT4. 
The complete data log is reported in Table~\ref{tab:obslog}.
The data on the fringe tracker (FT) detector were recorded at low spectral resolution (${\rm R}$~$\sim$~23) with a DIT of 0.85\,ms and those of the science (SC) detector at high spectral resolution 
(HR; ${\rm R}$~$\sim$~4000, i.e. $\Delta \varv \sim$70\,\kms). 
The three datasets were reduced using the GRAVITY pipeline~\citep[v1.2.1;][]{DRS}. HD\,31464 and HD\,37491 calibrators were used to retrieve the atmospheric transfer function.
The IRS\,2 spectrum was obtained by averaging the four HR UT spectra recorded in the three datasets. Standard telluric correction was also applied to the spectrum using HD\,31464 (SpT K0\,III) as a telluric standard star. 
The IRS\,2 spectrum was then flux calibrated adopting the 2MASS catalogue value. 
The spectral wavelength calibration was refined using several telluric absorption lines present along the spectrum. An average shift of $\sim$4.5\AA\ was applied.
To convert the observed wavelengths into radial velocities, we adopted a local standard of rest (lsr) velocity of 6\,km\,s$^{-1}$~\citep[][]{lenorzer}.

\section{Results}
\label{sec:res}
\begin{figure}[t]
        \centering
        \includegraphics[width=\columnwidth]{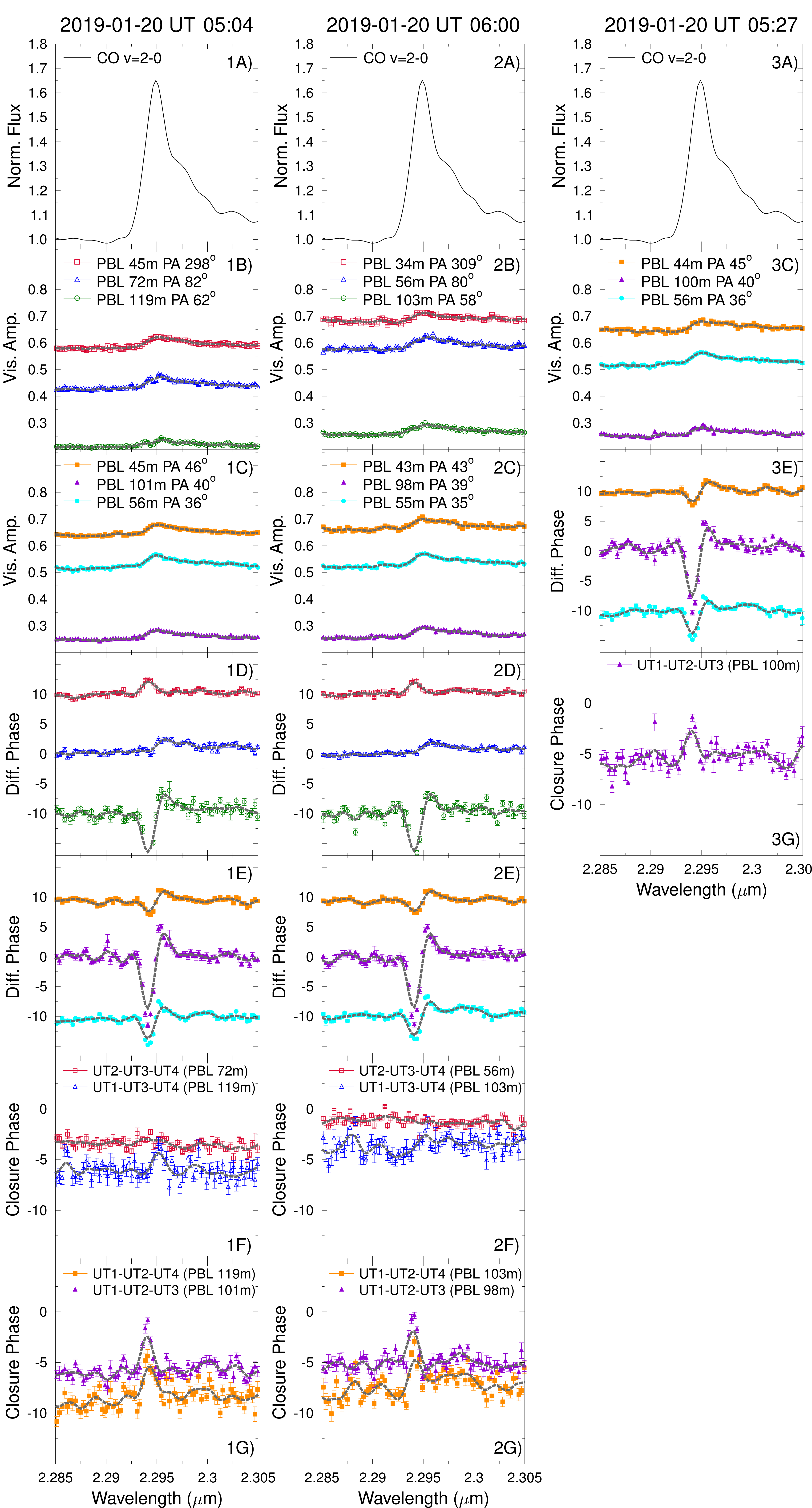}
    \caption{{\it Left: Panel\,1.} Interferometric measurements of the CO $v=2-0$ bandhead in NGC\,2024\,IRS2 for run\,1 (inserts 1A--1G).
                From  top to bottom: Total flux normalised to continuum (1A); wavelength-dependent visibilities for UT 3-4, 2-4, 1-4 (1B) and for UT 2-3, 1-3, 1-2 (1C); differential phases for UT 3-4,2-4,1-4 (1D) and
                for UT 2-3, 1-3, 1-2 (1E); and closure phases for UT 2-3-4, 1-3-4 triplets (1F) 1-2-4 and 1-2-3 triplets (1G).
                {\it Middle: Panel\,2.} Interferometric measurements of the CO $v=2-0$ and $v=3-1$ bandheads in NGC\,2024\,IRS2 for run\,3 (inserts 2A--2G). 
                {\it Right: Panel\,3.} Interferometric measurements of the CO $v=2-0$ and $v=3-1$ bandheads in NGC\,2024\,IRS2 for run\,2 (inserts 3A--3G).
                From  top to bottom: Total flux normalised to continuum (3A); wavelength-dependent visibilities for  UT 2-3, 1-3, 1-2 (3C); differential phases for UT 2-3, 1-3, 1-2 (3E); and closure phase for UT 1-2-3 (3G).
                For clarity, the differential phases of the first and last baselines are shifted by +10$\degr$ and -10$\degr$, respectively.}
    \label{fig:observables}
\end{figure}

Our GRAVITY/VLTI datasets provide us with the $K$-band spectrum of IRS\,2, six (or three for the second dataset taken with 3 UTs) spectrally dispersed visibilities ($V$) 
and differential phases ($DP$), and four (or one for the second dataset) closure phases ($CP$; see Fig.~\ref{fig:observables}), with rms uncertainties of $\sim$1\% for $V$, $\sim$1$\degr$ for $DP,$ 
and $\sim$2$\degr$ for $CP$.

The IRS\,2 spectrum displays a rising continuum with a bright Br$\gamma$ (2.166\,$\mu$m) line and four overtone CO bandheads (from $v=2-0$ to $v=5-3$, i.e. from 2.29 to 2.39\,$\mu$m) in emission.
No other lines are detected in the spectrum above a threshold of  three sigma.  
The continuum emission is resolved in all the baselines and is slightly asymmetric, displaying small closure phases ($\leq$8$\degr$) for the triangles with long baselines and closure phases
consistent with zero for the shortest baseline triangles ($\leq$60\,m). Spectrally dispersed visibilities and $DP$ and $CP$ signatures are detected in both Br$\gamma$ and CO lines.
Notably, the small continuum asymmetry also affects the observed $DP$ and $CP$ of the lines, causing the redshifted $DP$ peak to be systematically smaller than the blueshifted one and
making the $CP$ value at the line peak smaller than that of the continuum.

In this letter we report on the interferometric signatures of the CO bandheads, detected and analysed for the first time in a HMYSO, leaving the Br$\gamma$ line analysis to a forthcoming publication.
Figure~\ref{fig:observables} shows the interferometric observables (line profiles - inserts A; visibilities - inserts B \& C; differential phases - inserts D and E; closure phases
- inserts F and G) of the first CO bandhead ($v=2-0$) and adjacent continuum for the three runs (Panel\,1, 2 and 3). The interferometric observables of the other three CO bandheads ($v=3-1$, $v=4-2$ and $v=5-3$), which are
basically identical to the first one but slightly more noisy, are shown in the appendix (Figures~\ref{fig:observables2b}-\ref{fig:observables2d}).
Visibility values, $V$, around each bandhead peak are larger than the continuum visibilities at all the six baselines in the three runs
(see inserts B and C  in Fig.~\ref{fig:observables} and Figs~\ref{fig:observables2b}-\ref{fig:observables2d}), indicating that the CO-emitting region, though spatially resolved, is more compact than the continuum. 
$DP$ at four of the six baselines (or at all the three baselines of the second dataset) display an asymmetric `S' shape with values up to 12$\degr$ and small ($\geq$2$\degr$) values at the short baselines as well as 
at the intermediate baselines with $PA$ around 80$\degr$, indicating a clockwise rotating disc with a major axis $PA$ close to 170$\degr$.

\subsection{Continuum-emitting region}
\label{sec:continuum}
To estimate the size, inclination ($i$), and position angle of the dusty disc, we fit the continuum visibilities recorded with the GRAVITY FT using a simple geometric model, which assumes a point-like star and a resolved circumstellar disc, 
as in \cite{lazareff}. No additional extended halo component is required to fit the data, so we do not include it in the model. This is likely because the IRS\,2 outflow cavity walls (the main source of such an extended halo in HMYSOs) 
are located well beyond the UT field of view (FoV); $>$200\,mas vs. $\sim$40\,mas.
Therefore, the complex visibility ($V$) at spatial frequency ($u,v$) and at a certain wavelength ($\lambda$) consists of two components: $V(u,v,\lambda)=f_s(\lambda)+f_d(\lambda)V_d(u,v,\lambda)$, where $f_s$ and $f_d$ are the stellar and
disc contributions to the continuum flux ($f_s + f_d=1$), $V_d$ is the disc visibility, and the stellar visibility is assumed to be 1, as the star is not resolved at our spatial resolution ($\sim$0.17\,mas is the expected diameter 
of a B0 zero age main sequence - ZAMS - star at 423\,pc).  
A stellar contribution factor to the continuum flux of $f_s=0.07\pm0.04$ was first estimated, assuming a stellar spectral type of B0 and $A_V$ of 24\,mag~\citep{lenorzer}. The $f_s$ value, together with its uncertainty, 
is used as a starting value in the interferometric fit, but is kept as a free parameter during the visibility fitting process. 

We use the fitting tool described in \cite{lazareff} to test different models in their ability  to fit both $V^2$ and $CP$. We test ellipsoids and rings with Gaussian and non-Gaussian radial brightness distributions. 
The free parameters for the ellipsoid models are the flux contributions of $f_s$ and $f_d$, the flattening as $\cos\,i$, $PA$, the weighting for the radial brightness distribution $Lor$, 
which varies from a purely Gaussian to a purely Lorentzian distribution, and the half-flux semi-major axis $a$. 
The non-Gaussian models lead to $\chi_r^2$ values closer to 1. The ellipsoid and ring models converge towards the same set of parameters. 
Our best fit ellipsoid model ($\chi_r^2$=0.33) provides an inclination of 34$\degr\pm$1$\degr$, a $PA$ of about 166$\degr\pm$1$\degr$, and a disc diameter of 3.99$\pm$0.09\,mas (1.69$\pm$0.04\,au) as reported in Table~\ref{tab:fits} 
(see Figure~\ref{fig:continuum} in the Appendix for the continuum fit and Table~\ref{tab:continuum_fits} in the Appendix for the whole set of modelled parameters).
Notably, the derived $i$ value is equal to that inferred by \cite{chandler} (33$\degr$), who fit the $v=2-0$ CO bandhead (observed at high-spectral resolution $\mathcal R \sim 15\,000$) with a disc in Keplerian rotation. 
The disc contribution dominates the continuum emission in the $K$-band with a flux contribution of about 91\%, in full agreement with the estimate derived from the spectral type. In addition, we are able to correctly model the small 
closure phase signals with a skewed rim, originating from a pure inclination effect, with its maximum brightness roughly located westwards.

\subsection{CO continuum-subtracted visibilities and closure phases}
\label{sec:COpure}
Both size and geometry of the CO-emitting region can be determined from the pure (or continuum compensated) CO bandhead visibilities ($V_{CO}$). These are
estimated by subtracting the continuum contribution to the total line visibilities and by taking into account the line photocentre shifts~\citep[][]{weigelt07}:

\begin{equation}
 V_{CO} = \frac{\sqrt{\textbar F_{tot} V_{tot} \textbar^2 + \textbar F_{cont} V_{cont} \textbar^2 - 2 F_{tot} V_{tot}F_{cont} V_{cont} \cos \phi} }{F_{line}}, 
\end{equation}
where $F_{tot} = F_{cont} + F_{line}$, $F_{line}$, and $F_{cont}$ are the total, line, and continuum fluxes, respectively; $V_{cont}$ and $V_{tot}$ are the measured continuum and total visibilities, respectively; and $\phi$ is the differential phase.
We compute $V_{CO}$ for the four bandheads, in the three spectral channels around the bandhead peak, namely those with line-to-continuum ratio larger than 30\%.
Errors are estimated taking into account the uncertainties on the continuum and line fluxes, on the total visibilities, and on the differential phases for each 
spectral channel. Within the error bars, the three values are the same, and therefore we average the results obtaining less noisy values with average errors ($\Delta V_{CO}$) of 0.03. 
$V_{CO}$ and $\Delta V_{CO}$ per baseline and per bandhead are reported in Table~\ref{tab:CO_VIS}, along with the $(u,v)$ values, projected baseline length, $PA,$ and the UT telescope configuration.
To determine the size, inclination, and position angle of the CO-emitting region, we then fit the $V_{CO}$ datapoints in the $(u,v)$ plane using a bidimensional Gaussian with $FWHM_{CO}$, $i_{CO}$ and $PA_{CO}$ as free parameters. 
To derive the best values we use our own Python program based on the Monte–Carlo and Markov chain (MCMC) code $emcee$~\citep[][see also details on the method in Sect.~\ref{sec:appendix_CO_MCMC}]{emcee}.
We first model the visibilities of each single bandhead, the fit of which provides, within the error bars, very similar results in size (see Tab.~\ref{tab:fits}).
As the different bandheads are excited at different temperatures, the latter suggests that the radial extent ($\Delta R$) of the CO-emitting region must be relatively small
($\Delta R / R \leq$20\%).

To improve our results, we fit the whole set of visibilities, assuming that the bandhead-emitting regions have the same size and geometry. 
The best model ($\chi_r^2$=1.34; see MCMC marginal posterior distributions in Fig.~\ref{fig:MCMC} in the appendix) is shown in Figure~\ref{fig:CO_2D_fit}, 
which reports the bidimensional Gaussian projected on the $(u,v)$ plane along with the observed visibilities and their uncertainties. 
The CO-emitting region has a diameter of 2.74$\pm^{0.08}_{0.07}$\,mas (1.16$\pm$0.03\,au), namely it is located in the inner gaseous disc within the dusty rim, with $i_{CO}$ (32$\degr\pm$3$\degr$) and $PA_{CO}$
(168$\degr\pm^{5\degr}_{4\degr}$), matching the geometry of the dusty disc (see Table~\ref{tab:fits}) and indicating that both discs are coplanar. Notably, the inferred CO radius value (0.58\,au)
is within the range of values (0.28-0.84\,au) estimated by \cite{chandler}.

\begin{table}[t]
\caption{Diameter, inclination, and position angle derived from the best fit of continuum and CO bandheads. 1\,$\sigma$ uncertainties are reported.}
\label{tab:fits}
\centering
\begin{tabular}{c c c c c}
\hline \hline
Continuum & diameter & diameter & $i$ & $PA$ \\ 
      & [mas]  & [au]  & [$\degr$] & [$\degr$] \\
      \hline\      
      & 3.99$\pm^{0.08}_{0.1}$ & 1.69$\pm^{0.03}_{0.04}$ & 34$\pm$1  & 166$\pm$1 \\ 
      \hline\hline
Bandhead &  & &  &  \\ 
      \hline
All       & 2.74$\pm^{0.08}_{0.07}$ & 1.16$\pm$0.03 & 32$\pm$3  & 168$\pm^{5}_{4}$ \\
v = 2--0  & 2.9$\pm^{0.1}_{0.2}$ & 1.21$\pm^{0.04}_{0.08}$ & 33$\pm^{5}_{8}$  & 159$\pm^{8}_{5}$ \\ 
v = 3--1  & 2.6$\pm$0.1          & 1.10$\pm$0.04           & 28$\pm^{6}_{7}$  & 177$\pm^{14}_{12}$ \\
v = 4--2  & 2.8$\pm$0.1          & 1.18$\pm$0.04           & 32$\pm^{4}_{5}$  & 169$\pm^{11}_{7}$ \\
v = 5--3  & 2.5$\pm$0.1          & 1.06$\pm$0.04           & 33$\pm^{4}_{6}$  & 187$\pm^{9}_{12}$ \\
\hline
\end{tabular}
\end{table}


To retrieve any asymmetry from the CO-emitting region, we remove the continuum contribution from the line closure phase of each UT triangle, obtaining the so-called
closure differential phase (CDP), which is the closure of the pure differential phases of the CO bandheads. We compute the CDP for the seven triangles available and for the 
the first four bandheads around their peaks (averaging three to five spectral channels), namely where the line-to-continuum ratio is larger than 1.3. As a result, the CDP of the CO-emitting region is 
$\sim$0$\degr$ (within the uncertainties, which range form $\sim$5$\degr$ to $\sim$10$\degr$ for the first and the fourth bandhead, respectively). This indicates that the CO-emitting region is symmetric around the central source, and its
small CP signatures (see panels G in Fig.~\ref{fig:observables} and Figs.~\ref{fig:observables2b}-\ref{fig:observables2d} in the Appendix) arise from the continuum asymmetry.



\begin{figure}[t]
        \centering
        \includegraphics[width=8.6cm]{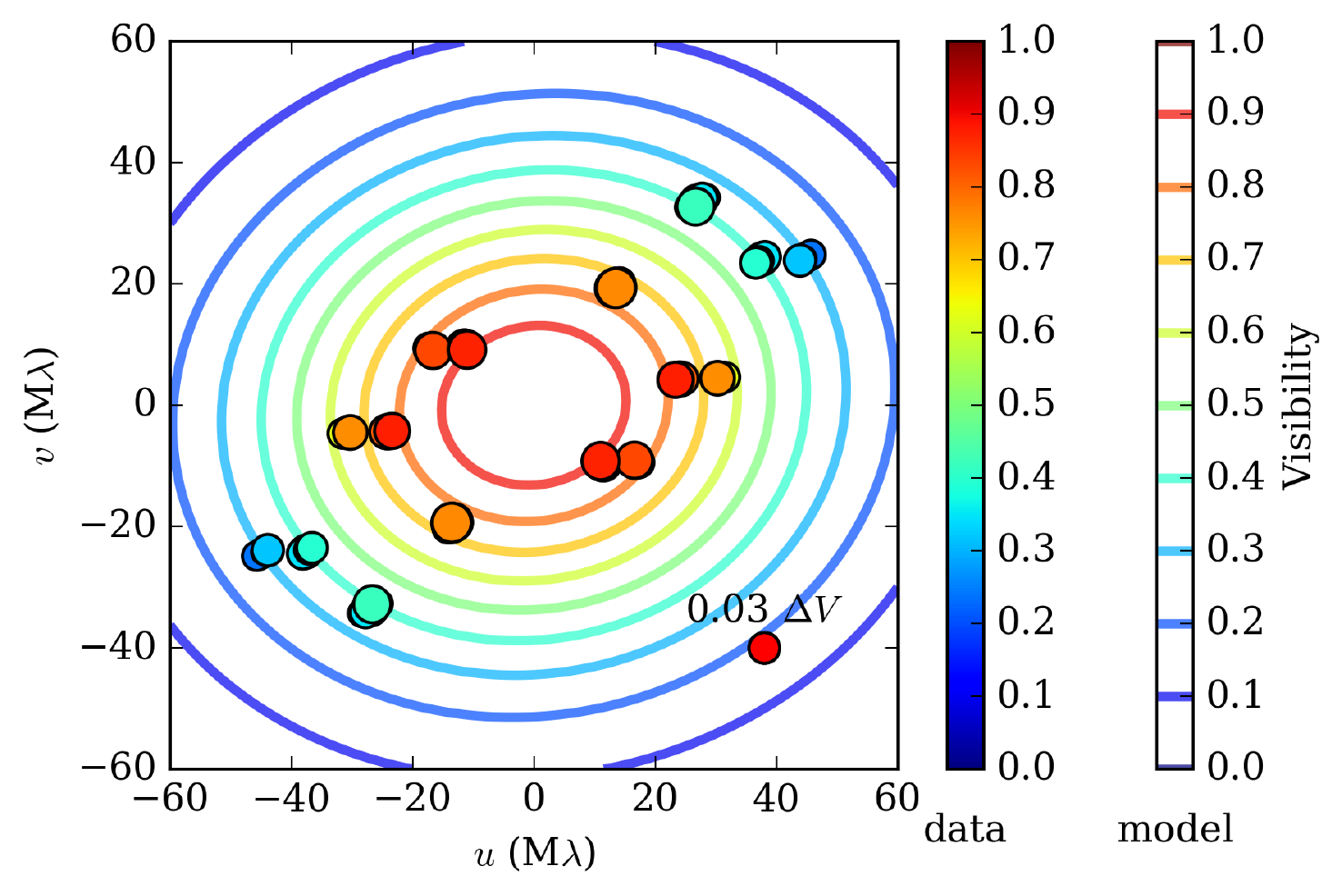}
        \caption{CO 2D Gaussian model (coloured ellipses) in the $(u,v)$ plane for the observed visibilities of the four bandheads (coloured circles). 
        Visibility values (from 0 to 1) of both model and observations are shown in scale from blue to red. Visibility uncertainties are represented with different diameters 
        (see red circle at the bottom right of the plot for an uncertainty average value of 0.03).
        }
            \label{fig:CO_2D_fit}
\end{figure}

\subsection{CO physical parameters}
\label{sec:physparam}
\begin{figure}[t]
        \centering
        \includegraphics[width=\columnwidth]{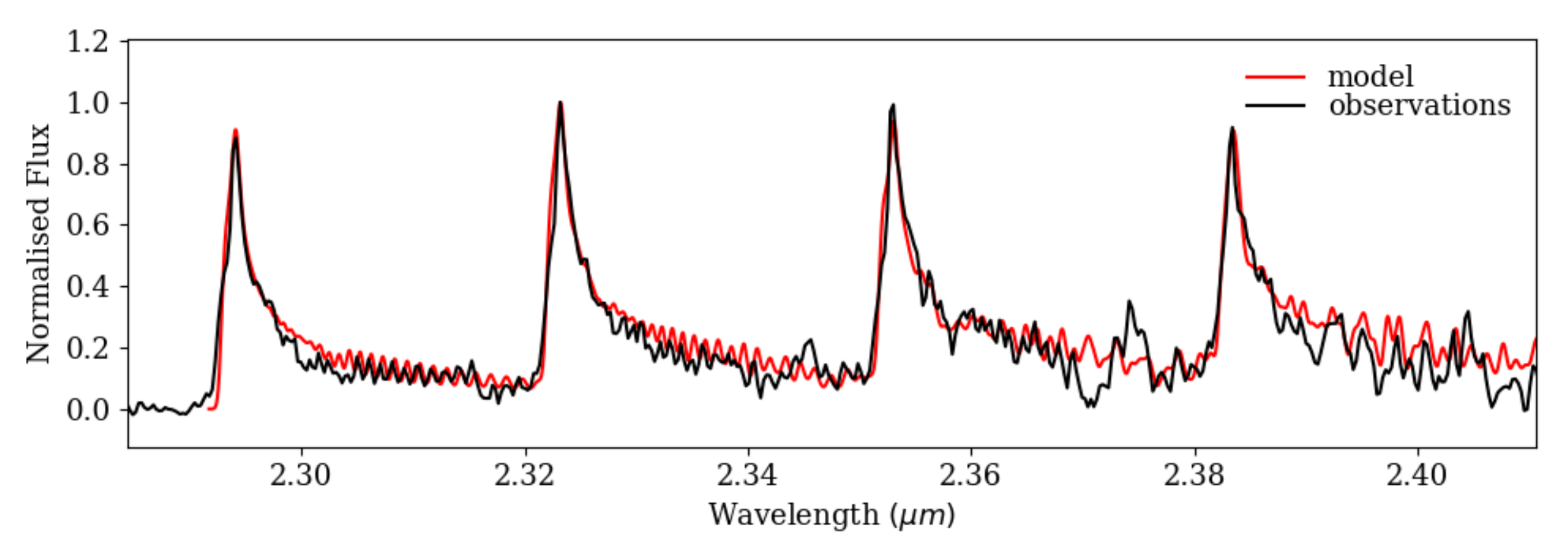}
        \caption{GRAVITY spectrum of the first four CO bandheads (in black) overplotted on our LTE model (in red). The GRAVITY spectrum is continuum subtracted and normalised to the peak of the second bandhead.}
            \label{fig:CO_fit}
\end{figure}

We use a CO local thermodynamic equilibrium (LTE) model~\citep[see][for a detailed description of model, code and error estimates]{koutoulaki} to derive the main physical parameters of the gas.
We model the CO bandhead profiles with a single ring in LTE with four free parameters: temperature $T_{CO}$, column density $N_{CO}$, turbulence velocity of the gas $\Delta v$, and projected Keplerian velocity 
($v_K \sin i$, where $i$ is the inclination of the disc plane with respect to the sky plane). 
A large grid of models was computed ranging over the free parameter space and then each resulting spectrum was convolved to the GRAVITY spectral resolution.
Our model is able to reproduce both the peaks and tails of the four bandheads  very well, although some portions of the tails are clearly missed due to the presence of strong telluric features.
Figure~\ref{fig:CO_fit} shows the spectrum of the first four CO bandheads (black curve) overplotted over our best model (red curve) with the following parameters: 
$T_{CO}$ = 2800$^{+300}_{-200}$\,K, $N_{CO}$ = (5$^{+5}_{-1}$)$\times$10$^{20}$\,cm$^{-2}$, 
$\Delta \varv$ = 1$^{+1}_{-0.5}$\,km\,s$^{-1}$, and $\varv_{K}\sin i$ = 84$^{+10}_{-20}$\,km\,s$^{-1}$. 
By measuring inclination and CO position from our interferometric data and the $\varv_{K}\sin i$ from the spectral fit, the stellar mass can be derived.
It is worth noting that such a measurement is not possible otherwise as the photospheric veiling in HMYSO is too high for a proper spectral-type estimate and
is more accurate than those derived with ALMA, for example, which include the whole disc mass. If we adopt the $i$ value from the dusty disc, we obtain an estimate of the Keplerian velocity
of $v_K \sim$150\,km\,s$^{-1}$, which at 0.58\,au implies a central mass of $M_*\sim$14.7$^{+2}_{-3.6}$\,M$_\sun$. This corresponds to a SpT=B0.5$\pm$0.5~\citep[i.\,e. $T_{eff}$ =29\,000$^{+2500}_{-3000}$\,K;][]{pecault}, assuming that IRS\,2 is on the ZAMS. 

\section{Origin of the CO bandheads in NGC\,2024\,IRS\,2}
\label{sec:discussion}
Our interferometric results demonstrate that the CO bandheads are emitted in the inner gaseous disc (located at 0.58\,au from the star), more specifically in a dust-free region well within the dust sublimation radius
(located at 0.85\,au from IRS\,2). We infer that IRS\,2 is a $\sim$15\,M$_\sun$ star with $T_{eff}\sim$29\,000\,K on ZAMS. As the inner gaseous region is free from dust grains, the CO molecules should be photodissociated by the stellar UV photons.
It is worth asking why the photo-dissociation does not happen and whether the CO-emitting region is in the disc midplane or more close to its surface. 
To answer the first question, we note that the IRS\,2 mass accretion rate ($\dot{M}_{acc}$) is $\sim$5$\times$10$^{-7}$\,M$_\sun$\,yr$^{-1}$~\citep{chandler,lenorzer}, 
which is high enough to make the gaseous disc optically thick~\citep[see][and references therein]{dullemond}. Therefore, the very inner gaseous disc should be able to shield the CO-emitting gas.
Moreover, the observed CO column density ($\sim$5$\times$10$^{20}$\,cm$^{-2}$) is much higher
than the value needed for the CO molecules to self-shield~\citep[$\sim$10$^{15}$\,cm\,$^{-2}$; see][]{ewine88,bik04}.
To answer the second question, we consider an optically thick Shakura-Sunyaev-type accretion disc around a B0.5 star on the ZAMS (with parameters 
$M_*$=14.7\,M$_\sun$, $R_*$=7\,R$_\sun$, $T_{eff}$=29\,000\,K, $\dot{M}_{acc}$=5$\times$10$^{-7}$\,M$_\sun$\,yr$^{-1}$). 
At 0.58\,au (i.e. where the CO is located), we infer surface density ($\Sigma$) values ranging from 10$^3$ to 10$^4$\,g\,cm$^{-2}$~\citep[using Equation~16 of][and varying the turbulent viscosity coefficient $\alpha$ from 0.1 to 0.01]{dullemond}.
A midplane temperature of $\sim$2000\,K can also be estimated~\citep[using Equation~15 of][]{dullemond}, assuming that the disc is heated by stellar radiation rather than viscous accretion, 
which becomes relevant at much higher mass accretion rates for HMYSOs~\citep[see e.g.][]{fedriani20}. In comparison, the total column density traced by the CO (assuming a factor of 10$^4$ between the CO and the total gas column density) 
is 8$^{+8}_{-2}$\,g\,cm$^{-2}$, which is three or four orders of magnitude lower than what is predicted for the disc midplane. In addition, the observed CO temperature ($T_{CO}$=2800$^{+300}_{-200}$\,K) is higher than that predicted
for the disc midplane but lower than what is expected on the disc surface ($\sim$4500\,K). At this temperature, the molecule is completely destroyed~\citep[][]{bosman}.
This reasoning indicates that the CO-emitting region is located well above the midplane but below the disc surface.


\section{Conclusions}
\label{sec:conclusion}
The main results of the first spectro-interferometric observations of the CO bandheads in the HMYSO NGC\,2024\,IRS\,2 are the following.
 
\begin{enumerate}
 \item The CO overtone (located at 0.58$\pm$0.02\,au from the star) is emitted from a dust-free region in the inner gaseous disc, which is coplanar with the inner dusty disc (located at 0.85$\pm$0.02\,au, with
 $i$=34$\degr\pm$1$\degr$ and $PA$=166$\degr\pm$1$\degr$) and of relatively small radial extent ($\Delta R / R \leq$20\%). The emitting region is located well above the disc midplane but below the disc surface.
 \item By modelling the CO spectrum, the following physical parameters are inferred: $T_{CO}$ = 2800$^{+300}_{-200}$\,K, $N_{CO}$ = (5$^{+5}_{-1}$)$\times$10$^{20}$\,cm$^{-2}$, $\Delta \varv$ = 1$^{+1}_{-0.5}$\,km\,s$^{-1}$, and $\varv_{K}\sin i$ = 84$^{+10}_{-20}$\,km\,s$^{-1}$. 
 \item By combining inclination and CO position from our interferometric data and the $\varv_{K}\sin i$ from the spectral fit, we provide a direct measurement of the stellar mass ($M_*\sim$14.7$^{+2}_{-3.6}$\,M$_\sun$) in a HMYSO.
\end{enumerate}

\begin{acknowledgements}
This paper is based on observations made with ESO Telescopes at the La Silla Paranal Observatory under programme IDs 0102.C-0408(C). We thank the technical, administrative, and scientific staff of the 
participating institutes and the observatory for their extraordinary support during the development, installation, and commissioning of GRAVITY.
A.C.G. wishes to thank Antonella Natta for her useful insights and comments. 
A.C.G. and T.P.R. have received funding from the European Research Council (ERC) under the European Union's Horizon 2020 research and innovation programme (grant agreement No.\ 743029).
R.F. acknowledges support from Science Foundation Ireland (grant 13/ERC/12907) and from a Chalmers Initiative on Cosmic Origins (CICO) postdoctoral fellowship. 
M.K. is funded by the Irish Research Council (IRC), grant GOIPG/2016/769 and Science Foundation Ireland (grant 13/ERC/12907).
R.G.L has received funding from Science Foundation Ireland under Grant No. 18/SIRG/5597.
K.P. acknowledges the funding of the French National Program of Stellar Physics (PNPS) and the grant from LabEx OSUG@2020 (Investissements d’avenir – ANR10LABX56),
A.A., P.G., P.G. were supported by Fundac\~{a}o para a Ci\^encia e a Tecnologia, with grants reference UID/FIS/00099/2013  and  SFRH/BSAB/142940/2018.
This research has made use of the Jean-Marie Mariotti Center \texttt{Aspro} and \texttt{SearchCal} services, \footnote{Available at http://www.jmmc.fr/} and 
of CDS Astronomical Databases SIMBAD and VIZIER \footnote{Available at http://cdsweb.u-strasbg.fr/}.
\end{acknowledgements}

\bibliographystyle{aa} 
\bibliography{references} 


\begin{appendix}
\label{sec:appendix}

\section{CO interferometric observables}
\label{sec:appendix_CO}

\begin{figure}[t]
        \centering
        \includegraphics[width=\columnwidth]{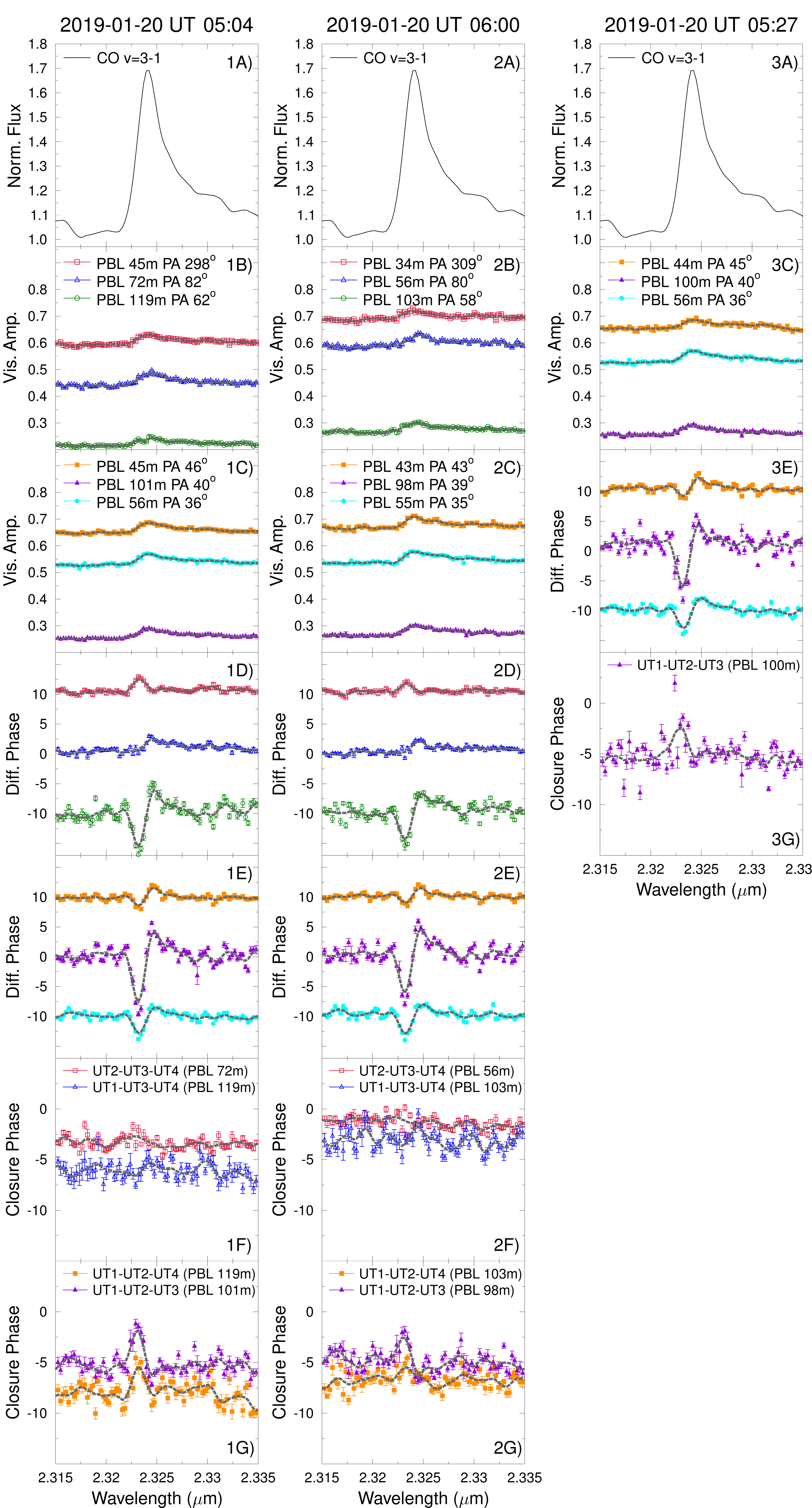}
    \caption{{\it Left: Panel\,1.} Interferometric measurements of the CO $v=3-1$ bandhead in NGC\,2024\,IRS2 for run\,1 (inserts 1A--1G).
                From  top to bottom: Total flux normalised to continuum (1A); wavelength-dependent visibilities for UT 3-4, 2-4, 1-4 (1B) and for UT 2-3, 1-3, 1-2 (1C); differential phases for UT 3-4,2-4,1-4 (1D) and
                for UT 2-3, 1-3, 1-2 (1E); closure phases for UT 2-3-4, 1-3-4 triplets (1F) 1-2-4 and 1-2-3 triplets (1G).
                {\it Middle: Panel\,2.} Interferometric measurements of the CO $v=4-2$ and $v=5-3$ bandheads in NGC\,2024\,IRS2 for run\,3 (inserts 2A--2G). 
                {\it Right: Panel\,3.} Interferometric measurements of the CO $v=4-2$ and $v=5-3$ bandheads in NGC\,2024\,IRS2 for run\,2 (inserts 3A--3G).
                From  top to bottom: Total flux normalised to continuum (3A); wavelength-dependent visibilities for  UT 2-3, 1-3, 1-2 (3C); differential phases for UT 2-3, 1-3, 1-2 (3E); and closure phase for UT 1-2-3 (3G).
                For clarity, the differential phases of the first and last baselines are shifted by +10$\degr$ and -10$\degr$, respectively.}
    \label{fig:observables2b}
\end{figure}

\begin{figure}[t]
        \centering
        \includegraphics[width=\columnwidth]{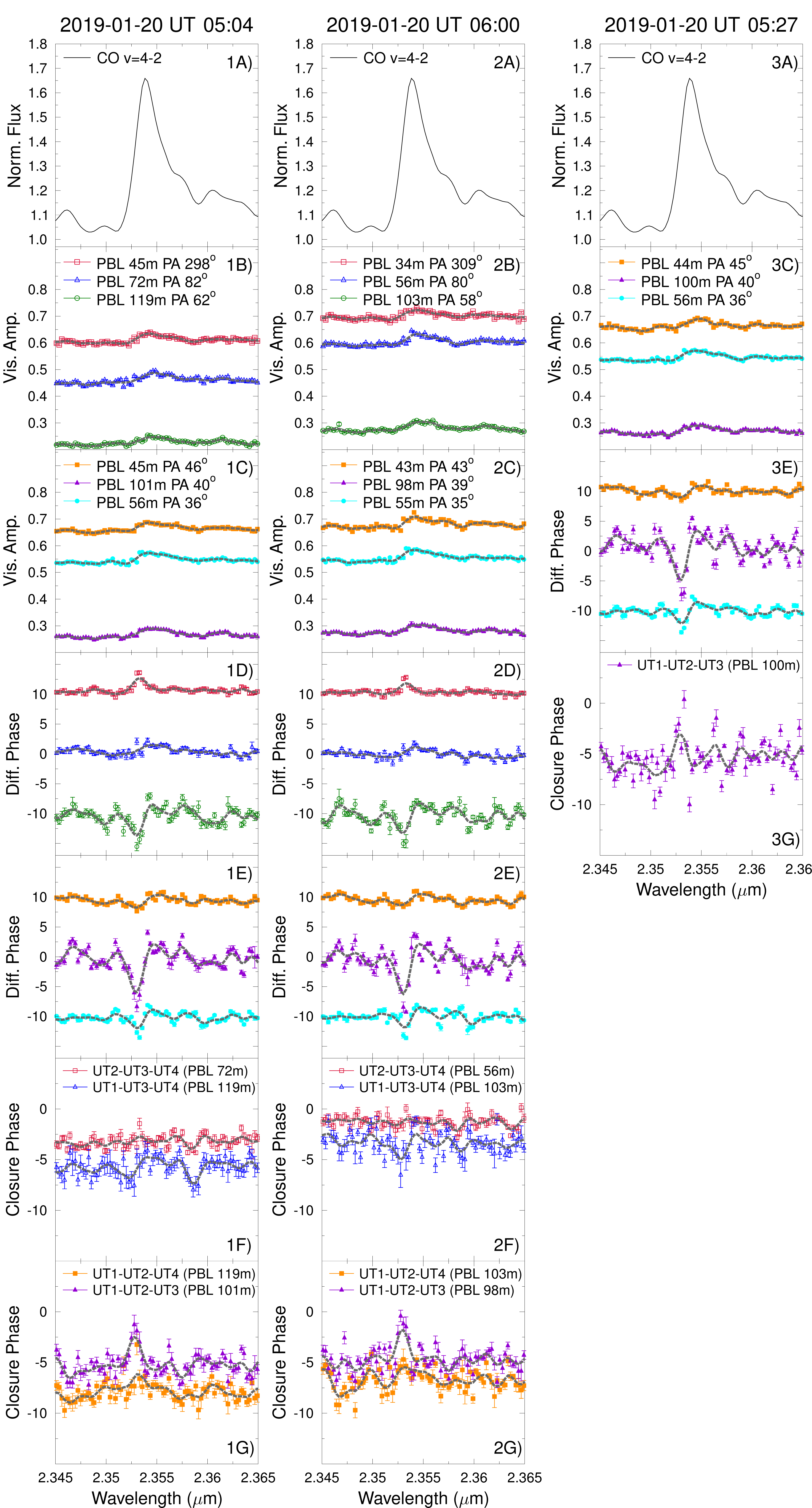}
    \caption{Same as Fig.~\ref{fig:observables2b} but for the CO $v=4-2$ bandhead }
    \label{fig:observables2c}
\end{figure}

\begin{figure}[t]
        \centering
        \includegraphics[width=\columnwidth]{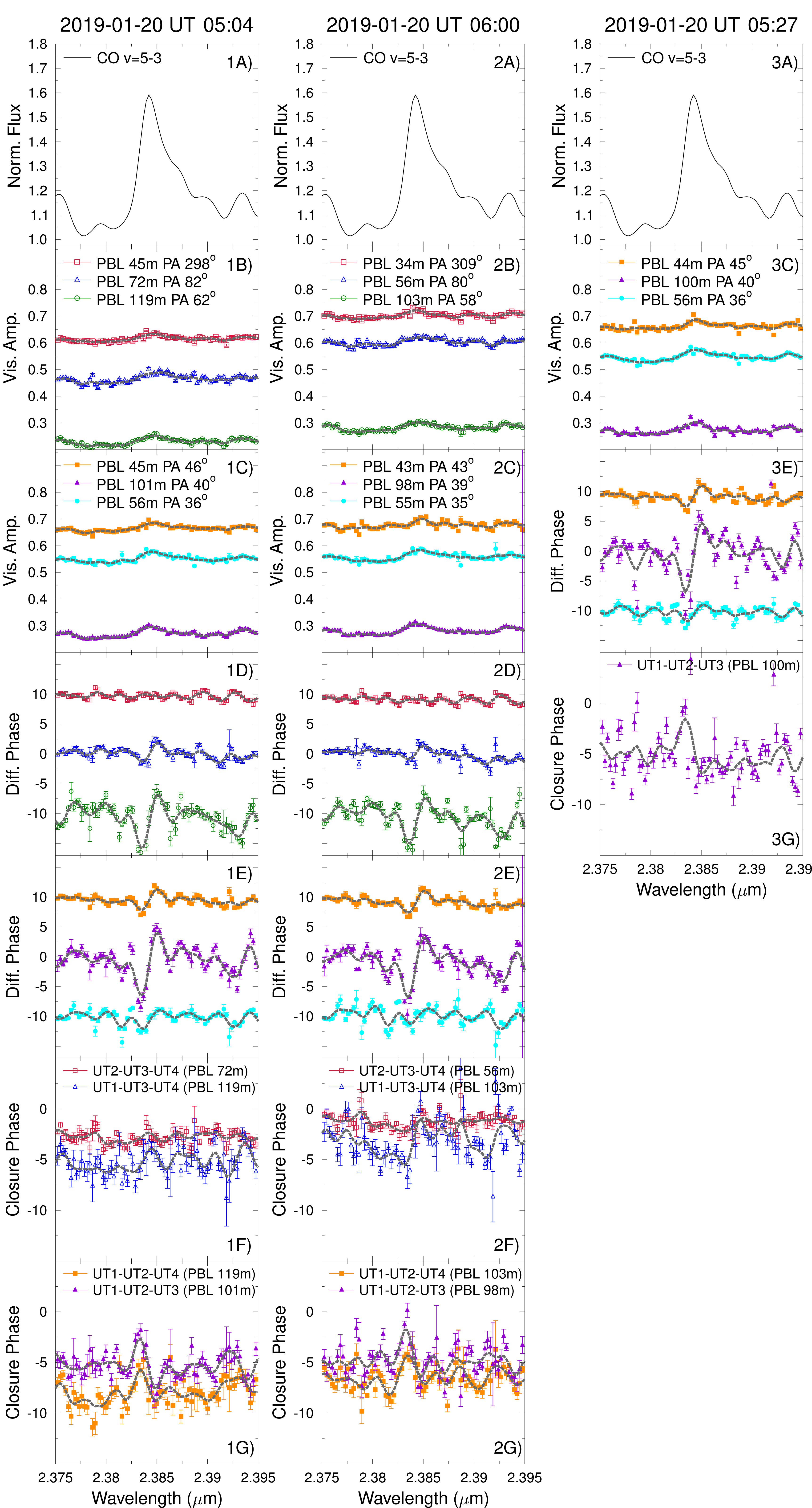}
    \caption{Same as Fig.~\ref{fig:observables2b} but for the CO $v=5-3$ bandhead }
    \label{fig:observables2d}
\end{figure}

Interferometric measurements of the CO $v=3-1$, $v=4-2$ and $v=5-3$ bandheads are reported in Figures~\ref{fig:observables2b}, \ref{fig:observables2c}, and \ref{fig:observables2d}, respectively.

\section{Continuum fit}
\label{sec:appendix_continuum}
\begin{figure}[t]
        \centering
        \includegraphics[width=\columnwidth]{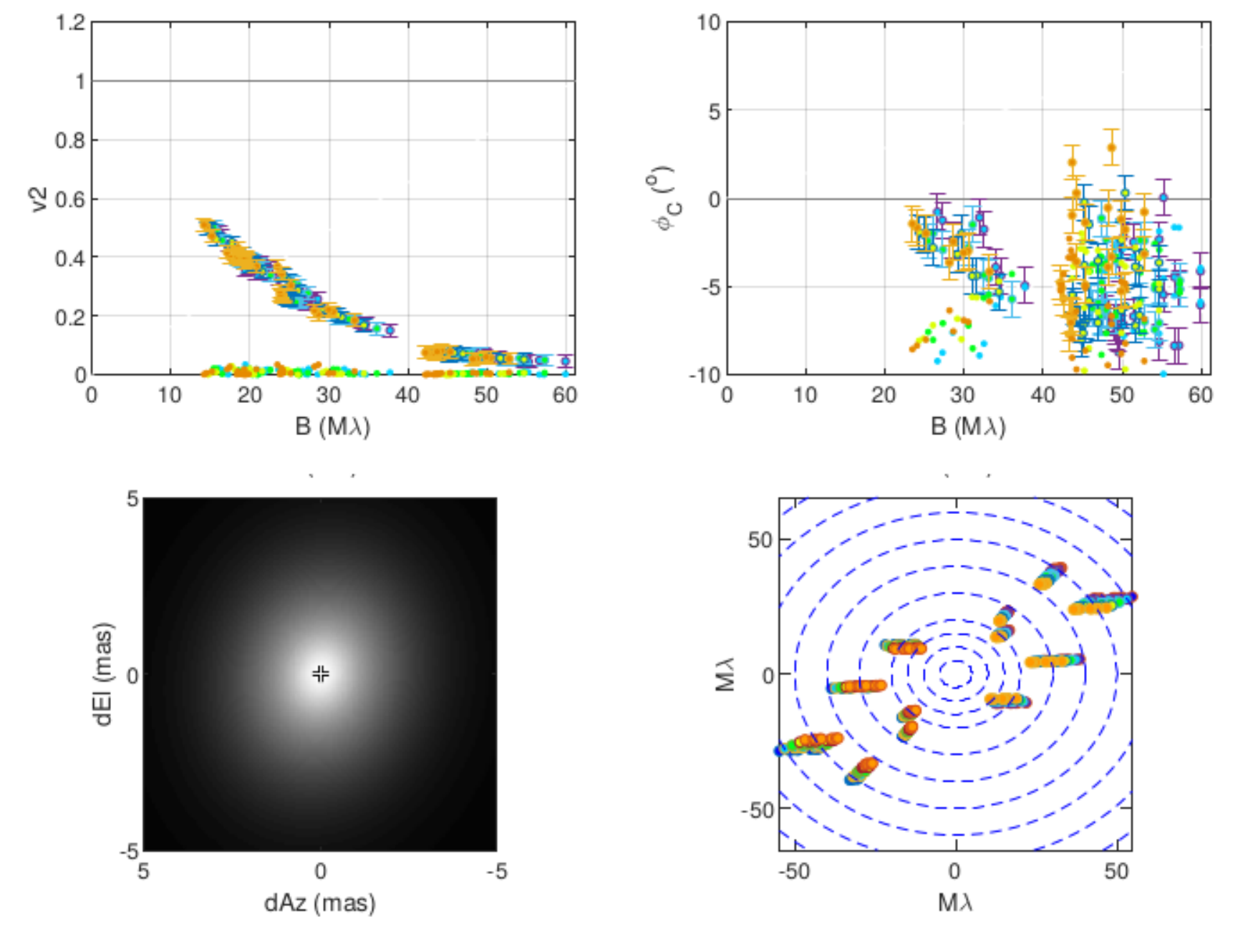}
    \caption{Continuum model results. Continuum visibilities as a function of wavelength along with the best continuum fit in the $(u,v)$ plane.
    For the $V^2$ and $CP$ plots, the absolute value of the fit residuals is shown by dots at the bottom of the plot.
 {\it Upper left panel.} Visibility squared $V^2$ vs. projected baseline in units of M$\lambda$.
    {\it Upper right panel.} $CP$ vs. the largest projected baseline of each triangle.
    {\it Lower left panel.} Halftone image of the circumstellar component resulting from a non-Gaussian Ellipsoid fit with m = 1 azimuthal modulation. The stellar position is shown by a cross symbol.
    {\it Lower right panel.} $(u,v)$ plane of the observations. }
    \label{fig:continuum}
\end{figure}

Continuum model parameters are reported in Table~\ref{tab:continuum_fits} and results are shown in Fig.~\ref{fig:continuum}.

\begin{table}[t]
\caption{Parameters derived from the best fit of the continuum. }
\label{tab:continuum_fits}
\centering
\begin{tabular}{c c }
\hline \hline
Parameter & value \\ 
     \hline
Diameter  &  3.99$\pm^{0.08}_{0.1}$\,mas \\
$i$  & 34$\degr\pm$1$\degr$   \\
$PA$  & 166$\degr\pm$1$\degr$  \\
$f_s$ & 0.09 \\
$f_d$ & 0.91 \\
$f_{lor}$\tablefootmark{a} & 0.58$\pm$0.02 \\
\hline
\end{tabular}
\tablefoot{
\tablefoottext{a}{Weighting for radial distribution and ranges between 0 (for a Gaussian radial distribution) and 1 (for a Lorentzian radial distribution).}
}
\end{table}

\section{MCMC approach for fitting the CO pure line visibilities}
\label{sec:appendix_CO_MCMC}

\begin{figure*}[t]
        \centering
        \includegraphics[width=\columnwidth]{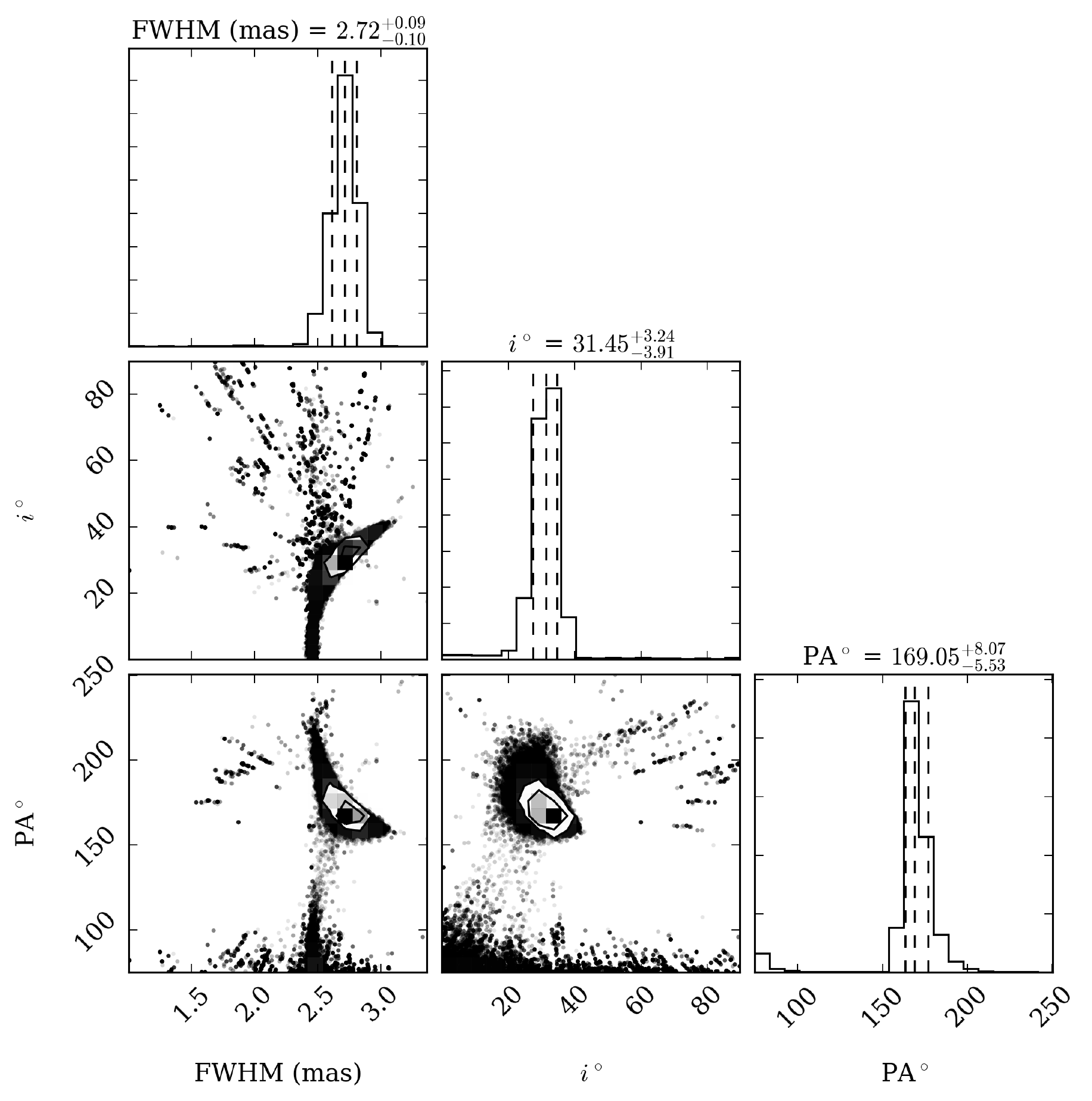}\includegraphics[width=\columnwidth]{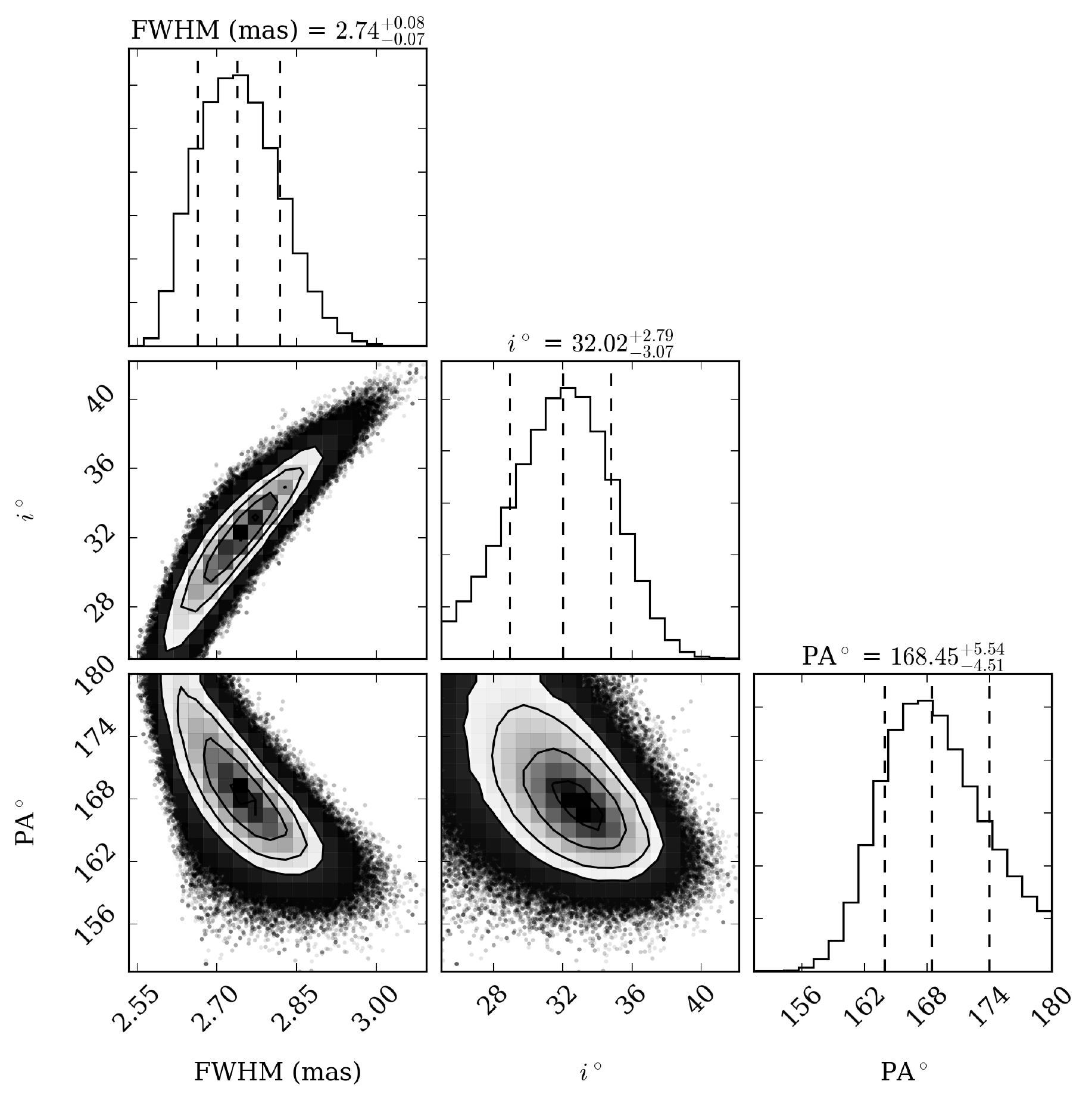}
         \caption{{\it Left Panel.} Marginal posterior distribution of the CO model for the $FWHM$, $i$, and $PA$, exploring the full range of space parameters (see text). We note that the $PA$ varies from
         75$\degr$ to 255$\degr$ for graphical reasons to keep the distribution in a single curve. The vertical dashed lines represent the 16$^{th}$, 50$^{th}$, 
         and 84$^{th}$ percentiles reported on the top of each marginal posterior distribution plot. {\it Right Panel.} Same as left panel but for a more constrained range of parameters.}
    \label{fig:MCMC}
\end{figure*}

To model the pure line visibilities of the CO-emitting region (see Table~\ref{tab:CO_VIS}), we assume a simple bi-dimensional Gaussian distribution in the $(u,v)$ plane:

\begin{equation}
    V = e ^{\left[\frac{\pi^2}{4\ln(2)}FWHM^2\left((u\sin PA+v\cos PA)^2+\cos^2i(v\sin PA - u\cos PA)^2\right)\right]}
     \label{eq:visibility}
,\end{equation}
where $FWHM$, $i$, and $PA$ are free parameters. We adopt a Bayesian MCMC approach to constrain the three free parameters of the model. We sample the parameter space using the $emcee$ 
module~\citep[see][for a detailed description of the method]{emcee}. 
We set the prior distribution to be uniform (i.e. non-informative prior) and the posterior distribution is therefore given by the product between the prior distribution function and the likelihood function
given by $\chi^2 = \sum (V_{obs} - V_{mod})^2/\sigma^2_V$ being $\sigma_V$ the variance of the data. 
We ran the MCMC with 1000 walkers and for 1000 steps. We set a burn-in period of 10$\%$, to account for the warm-up period of the chain. 
To avoid local minima, we first explore a large range of the space parameters (i.e. $FWHM\in[1,4\,mas]$, $i\in[0\degr,90\degr]$,$PA\in[0\degr,180\degr]$; see marginal posterior distributions in Figure~\ref{fig:MCMC}, left panel)
and then we reduce the range around the values obtained from those distributions (i.e. $FWHM\in[2.5,3.1\,mas]$, $i\in[25\degr,45\degr]$,$PA\in[150\degr,180\degr]$; see posterior distributions in 
Figure~\ref{fig:MCMC}, right panel). Assuming a Gaussian distribution, the most likely value is that of the 50$^{th}$ percentile, whereas the 1$\sigma$ uncertainty is given by the 
values falling in the 16$^{th}$ and 84$^{th}$ percentiles, represented in the histograms with vertical dashed lines. These values are given on top of each marginal posterior distribution plot.
We then calculate the $\chi^2_r$ for the most likely values.

\begin{table*}[t]
\caption{CO bandheads pure line visibilities}
\label{tab:CO_VIS}
\centering
\begin{tabular}{c c c c c c c c}
\hline \hline
$u$ & $v$ & $V_{CO}$ & $\Delta V_{CO}$ & PBL & PA & baseline & CO bandhead \\ 
 (M$\lambda$)  & (M$\lambda$)  &  &  & (m) & ($\degr$) & & (overtone)\\
      \hline
      \hline
-11.47   &    9.56  &   0.82 &  0.04  &   34   &  309 &   UT34  & v=2--0 \\
-12.85   &  -13.95  &   0.92 &  0.04  &   43   &  43  &   UT23  & v=2--0 \\
-13.64   &  -14.01  &   0.91 &  0.03  &   44   &  45  &   UT23  & v=2--0 \\
-14.22   &  -14.08  &   0.91 &  0.03  &   45   &  46  &   UT23  & v=2--0 \\
-17.29   &    9.42  &   0.76 &  0.03  &   45   &  298 &   UT34  & v=2--0 \\
-13.77   &  -20.07  &   0.74 &  0.03  &   55   &  35  &   UT12  & v=2--0 \\
-14.17   &  -20.20  &   0.72 &  0.03  &   56   &  36  &   UT12  & v=2--0 \\
-14.10   &  -20.13  &   0.71 &  0.03  &   56   &  36  &   UT12  & v=2--0 \\
-24.32   &  -4.39   &   0.81 &  0.04  &   56   &  80  &   UT24  & v=2--0 \\
-31.51   &  -4.66   &   0.62 &  0.03  &   72   &  82  &   UT24  & v=2--0 \\
-26.62   &  -34.02  &   0.37 &  0.03  &   98   &  39  &   UT13  & v=2--0 \\
-27.74   &  -34.14  &   0.35 &  0.03  &   100  &  40  &   UT13  & v=2--0 \\
-28.39   &  -34.28  &   0.35 &  0.02  &   101  &  40  &   UT13  & v=2--0 \\
-38.09   &  -24.46  &   0.35 &  0.03  &   103  &  58  &   UT14  & v=2--0 \\
-45.68   &  -24.86  &   0.24 &  0.03  &   119  &  62  &   UT14  & v=2--0 \\
-11.32   &    9.44  &   0.86 &  0.04  &   34   &  309 &   UT34  & v=3--1 \\
-12.69   &  -13.77  &   0.95 &  0.04  &   43   &  43  &   UT23  & v=3--1 \\
-13.46   &  -13.83  &   0.93 &  0.03  &   44   &  45  &   UT23  & v=3--1 \\
-14.04   &  -13.90  &   0.94 &  0.03  &   45   &  46  &   UT23  & v=3--1 \\
-17.06   &    9.30  &   0.80 &  0.03  &   45   &  298 &   UT34  & v=3--1 \\
-13.59   &  -19.81  &   0.78 &  0.04  &   55   &  35  &   UT12  & v=3--1 \\
-13.92   &  -19.87  &   0.74 &  0.04  &   56   &  36  &   UT12  & v=3--1 \\
-13.99   &  -19.94  &   0.75 &  0.04  &   56   &  36  &   UT12  & v=3--1 \\
-24.00   &  -4.33   &   0.83 &  0.04  &   56   &  80  &   UT24  & v=3--1 \\
-31.10   &  -4.60   &   0.69 &  0.03  &   72   &  82  &   UT24  & v=3--1 \\
-26.28   &  -33.58  &   0.40 &  0.03  &   98   &  39  &   UT13  & v=3--1 \\
-27.38   &  -33.70  &   0.38 &  0.03  &   100  &  40  &   UT13  & v=3--1 \\
-28.03   &  -33.84  &   0.38 &  0.02  &   101  &  40  &   UT13  & v=3--1 \\
-37.60   &  -24.14  &   0.37 &  0.04  &   103  &  58  &   UT14  & v=3--1 \\
-45.09   &  -24.54  &   0.27 &  0.03  &   119  &  62  &   UT14  & v=3--1 \\
-11.17   &    9.32  &   0.83 &  0.03  &   34   &  309 &   UT34  & v=4--2 \\
-12.52   &  -13.60  &   0.92 &  0.03  &   43   &  43  &   UT23  & v=4--2 \\
-13.29   &  -13.65  &   0.88 &  0.03  &   44   &  45  &   UT23  & v=4--2 \\
-13.86   &  -13.72  &   0.91 &  0.01  &   45   &  46  &   UT23  & v=4--2 \\
-16.85   &    9.18  &   0.79 &  0.03  &   45   &  298 &   UT34  & v=4--2 \\
-13.42   &  -19.56  &   0.77 &  0.03  &   55   &  35  &   UT12  & v=4--2 \\
-13.74   &  -19.62  &   0.71 &  0.03  &   56   &  36  &   UT12  & v=4--2 \\
-13.81   &  -19.68  &   0.74 &  0.03  &   56   &  36  &   UT12  & v=4--2 \\
-23.70   &  -4.28   &   0.84 &  0.03  &   56   &  80  &   UT24  & v=4--2 \\
-30.71   &  -4.54   &   0.66 &  0.03  &   72   &  82  &   UT24  & v=4--2 \\
-25.94   &  -33.15  &   0.39 &  0.02  &   98   &  39  &   UT13  & v=4--2 \\
-27.03   &  -33.27  &   0.36 &  0.03  &   100  &  40  &   UT13  & v=4--2 \\
-27.67   &  -33.40  &   0.36 &  0.01  &   101  &  40  &   UT13  & v=4--2 \\
-37.12   &  -23.83  &   0.36 &  0.03  &   103  &  58  &   UT14  & v=4--2 \\
-44.51   &  -24.22  &   0.27 &  0.02  &   119  &  62  &   UT14  & v=4--2 \\
-11.01   &  9.19    &   0.87 &  0.05  &   34   &  309 &   UT34  & v=5--3  \\
-12.34   &  -13.40  &   0.95 &  0.06  &   43   &  43  &   UT23  & v=5--3  \\
-13.10   &  -13.46  &   1.00 &  0.07  &   44   &  45  &   UT23  & v=5--3  \\
-16.60   &  9.05    &   0.84 &  0.04  &   45   &  298 &   UT34  & v=5--3  \\
-13.66   &  -13.52  &   1.00 &  0.04  &   45   &  46  &   UT23  & v=5--3  \\
-13.23   &  -19.28  &   0.82 &  0.05  &   55   &  35  &   UT12  & v=5--3  \\
-13.55   &  -19.34  &   0.76 &  0.05  &   56   &  36  &   UT12  & v=5--3  \\
-13.61   &  -19.40  &   0.80 &  0.05  &   56   &  36  &   UT12  & v=5--3  \\
-23.36   &  -4.21   &   0.88 &  0.04  &   56   &  80  &   UT24  & v=5--3  \\
-30.26   &  -4.48   &   0.76 &  0.04  &   72   &  82  &   UT24  & v=5--3  \\
-25.57   &  -32.68  &   0.44 &  0.02  &   98   &  39  &   UT13  & v=5--3  \\
-26.64   &  -32.79  &   0.42 &  0.05  &   100  &  40  &   UT13  & v=5--3  \\
-27.27   &  -32.92  &   0.42 &  0.03  &   101  &  40  &   UT13  & v=5--3  \\
-36.58   &  -23.49  &   0.40 &  0.03  &   103  &  58  &   UT14  & v=5--3  \\
-43.88   &  -23.88  &   0.32 &  0.03  &   119  &  62  &   UT14  & v=5--3  \\
\hline
\end{tabular}
\end{table*}

\section{Modelling results from continuum model plus CO geometric model}
\label{sec:appendix_CO_models}

\begin{figure*}[t]
        \centering
        \includegraphics[width=\columnwidth]{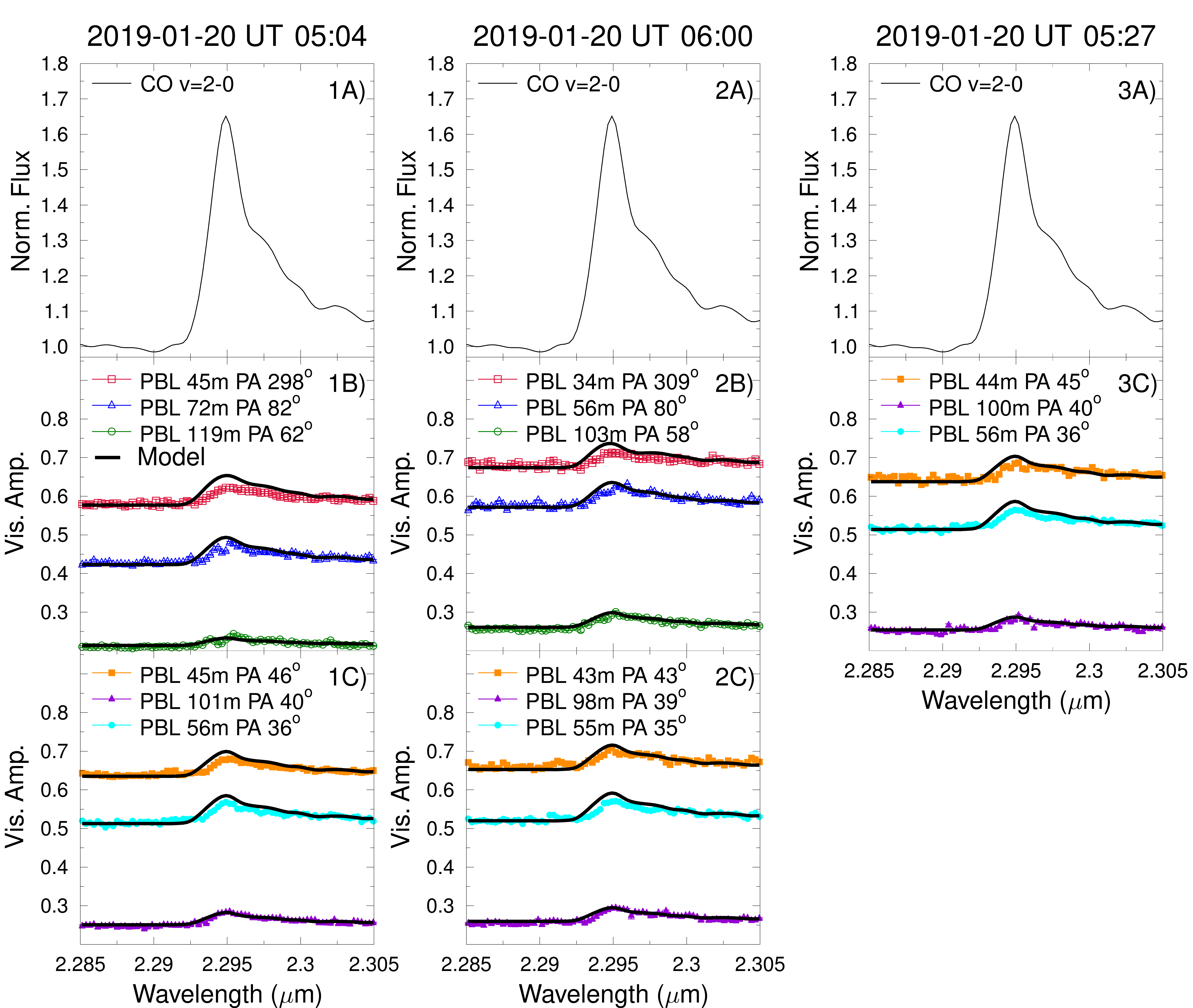}
         \caption{Observed visibilities (line plus continuum) of the CO $v=2-0$ bandhead vs. visibilities derived from our continuum model (see Sect.~\ref{sec:continuum} and Appendix~\ref{sec:appendix_continuum})
         plus the CO geometrical model from Eq.~\ref{eq:visibility} (see Sect.~\ref{sec:COpure} and Appendix~\ref{sec:appendix_CO_MCMC}). Model visibilities for each baseline and run are overplotted as black continuous lines.}
    \label{fig:model}
\end{figure*}

To verify the consistency of our modelling results, we checked the visibilities obtained from our continuum model (see Sect.~\ref{sec:continuum} and Appendix~\ref{sec:appendix_continuum})
plus the CO geometrical model obtained from the pure line visibilities (see Sect.~\ref{sec:COpure} and Appendix~\ref{sec:appendix_CO_MCMC}) against the observed visibilities.
We obtain similar results for the four observed bandheads and, as an example, Figure~\ref{fig:model} shows the results for the CO $v=2-0$ bandhead.  
Our model (black continuous line) perfectly matches the observed visibilities at the long baselines, whereas, at the short baselines, a small discrepancy (from 0.01 to 0.02) can be noted along the blueshifted shoulders and at peaks of the 
bandheads.

\end{appendix}

\end{document}